# Compactly packaged superconducting nanowire single-photon detector with an optical cavity for multichannel system

**Shigehito Miki,**[1,*] **Masanori Takeda,**[2] **Mikio Fujiwara,**[3] **Masahide Sasaki,**[3] **and Zhen Wang**[1]

[1]*Kansai Advanced Research Center, National Institute of Information and Communications Technology, 588-2, Iwaoka, Iwaoka-cho, Nishi-ku, Kobe, Hyogo 651-2492, Japan*
[2]*National Astronomical Observatory of Japan, 2-21-1 Osawa, Mitaka, Tokyo 181-8588, Japan*
[3]*National Institute of Information and Communications Technology, 4-2-1 Nukui-Kitamachi, Koganei, Tokyo 184-8795, Japan*
[*]*s-miki@nict.go.jp*

**Abstract:** We developed superconducting nanowire single-photon detectors with an optical cavity (OC-SNSPDs) for multichannel systems. For efficient coupling, the devices were installed in compact fiber-coupled packages after their substrate thickness was reduced from 400 to 45 μm. The measured detection efficiency (DE) measurement at different substrate thicknesses and the estimation of optical coupling efficiency indicated that ~98% of the input light beam could be radiated on a $15 \times 15$ μm$^2$ nanowire area from behind the substrate. The DEs of a NbN OC-SNSPD system were observed to be 9.5% and 25% at 1550 nm and 1310 nm, respectively (dark-count rate: 100 c/s).

**OCIS codes:** (040.5160) Photodetectors; (270.5570) Quantum detectors.

## 1. Introduction

Practical single-photon detectors are one of the most important components in the fields of quantum information and communications technology. In particular, the distribution length and secure key generation rate of quantum key distribution (QKD) systems are constrained by the performance of single-photon detectors. Recently, superconducting nanowire single-photon detectors (SNSPDs) have attracted in these fields because of their broadband sensitivity from the visible to the near-infrared wavelengths, excellent timing resolution, high counting rate, and low dark-count rate [1,2]. In particular, multichannel SNSPD systems based on the closed-cycle cryocooler are attractive from a practical perspective because they offer good performance and can operate continuously without any cryogen. They have been successfully employed in QKD experiments to boost both transmission distance and key generation rate [3–6]. Thus far, SNSPD systems used in QKD experiments have had a system detection efficiency (DE) of a few percent at a wavelength of 1550 nm and a dark-count rate of 100 c/s [7,8]. However, further improvement in the system performance, especially in the system DE, is highly desirable.

The system DE is determined by the product of the optical coupling efficiency between the incident light and the active area, the intrinsic photon-absorption coefficient of the superconducting nanowire, and the probability of electrical pulse generation after photon absorption. The first factor depends on the fiber coupling technique. The other two are nontrivial factors. The second factor depends on the optical properties of the nanowire material and the enclosing layers. The last factor is equal to the intrinsic DE of the nanowire and depends on the superconducting properties, geometrical structure, and homogeneity of the nanowire (i.e., thickness, width, and film quality).

This paper focuses on how the second factor, i.e., the intrinsic photon-absorption coefficient, can be increased by introducing an optical cavity in the SNSPD device. There are two approaches for configuring the SNSPD with an optical cavity (OC-SNSPD): (1) fabricating the nanowire on top of the optical cavity structure with front illumination and (2) integrating the optical cavity structure on top of the nanowire with rear illumination through the substrate. The first approach is convenient for achieving fiber-coupled packages, but the fabrication of high-quality uniform nanowires on top of an optical cavity is very difficult. The second approach can solve this difficulty because the nanowire can be fabricated directly on a substrate. However, one must pay special attention to ensure efficient optical coupling of the incident light to the active area of the nanowire through the substrate. This implies that the distance between the incident light exits and the active area of the device will be longer than the substrate thickness, causing an expansion of the incident light spot at the active area of the device. One possible solution to this problem is to use an appropriate lens with a working distance that is much longer than the substrate thickness and to adjust the position between the illuminated spot and the active area using a low-temperature nanostage [9]. However, this technique requires a large and complex system and is hence unsuitable for application to a multichannel system.

In this paper, we report on the fabrication of OC-SNSPD chips and the development of compact fiber-coupled packaging technique for multichannel GM cryocooler systems. In order to obtain high optical coupling efficiency in the package, we reduced the substrate thickness of the OC-SNSPD chips to 45 μm by using a mechanical polishing method. We verified the effectiveness of our packaging technique and the advantages of integrating optical cavity, and we evaluated the system DE performance of the OC-SNSPD.

## 2. Device design and fabrication

The OC-SNSPD consists of three components in the following order: a 100-nm-thick Au mirror, 250-nm-thick SiO cavity, and 4-nm-thick NbN nanowire; further, an antireflection (AR) layer is present underneath the substrate, as shown in Figure 1. The Au mirror and SiO cavity are designed to act as an optical cavity for a single photon at a wavelength of 1550 nm. The NbN nanowire and a coplanar waveguide line were first fabricated on a single-crystal MgO substrate. A method for growing ultrathin NbN epitaxial films and the fabrication process for nanowire devices have been described in [10]. The SiO and Au films were sequentially deposited by using the high vacuum evaporation method after patterning a square window on the active area of the nanowire by photoresist masking. Ion-beam cleaning was intentionally not performed prior to the deposition of SiO thin films, to prevent damage to the nanowire. Figure 2 (a) shows an optical micrograph of a fabricated SNSPD device with an optical cavity.

The thickness of the substrate was reduced to the desired thickness for effective optical coupling of the incident light to the active area by using a mechanical polishing system. After dicing the substrate to a size of $3 \times 3$ mm$^2$, the device was positioned facing downward and was fixed to a table in the polishing system; the rear of the substrate was pressed against a lapping sheet. The substrate was polished by rotating the lapping sheet at 120 rpm. The thickness of the substrate could be reduced at a rate of 3 μm/min. Since the thickness of the substrate was monitored by a micrometer, it was possible to control the thickness of the substrate with an accuracy within 5 μm. The optimization of the substrate thickness is described in section 3. As a typical example of this process, Figure 2 (b) shows a cross-sectional microphotograph of the substrate whose thickness was reduced to 45 μm. Although the substrate becomes mechanically weak with a decrease in the thickness, the 45-μm-thick substrate was still strong enough to treat and to carry out wire bonding.

As a necessary step after the reduction of the substrate thickness, the rear surface of the substrate was spin coated with a transparent fluoropolymer (refractive index: 1.34), which act as an AR layer. The thickness of the fluoropolymer was controlled to be approximately 250 nm by varying the spin-coating rotation speed, which corresponds to λ/4 at a wavelength of 1550 nm.

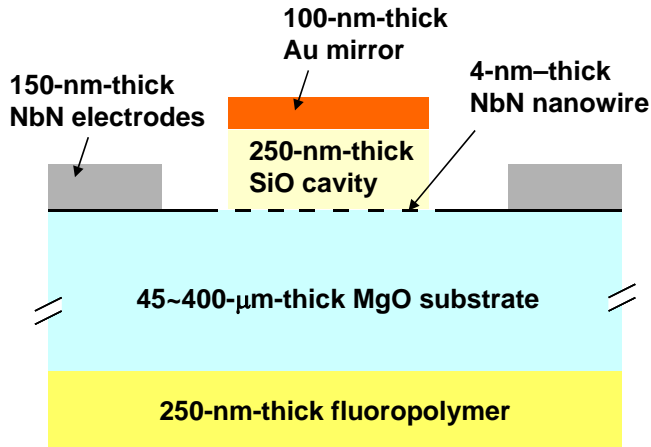

Fig. 1. Cross-sectional diagram of an OC-SNSPD chip.

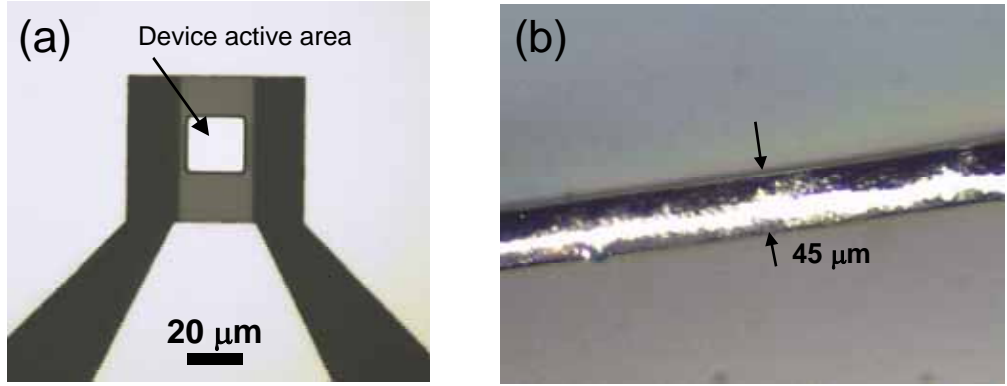

Fig. 2. (a) Optical micrograph of the nanowire area in the OC-SNSPD. (b) Cross-sectional micrograph of the substrate whose thickness was reduced to 45 μm.

**3. Device packaging and measurement setup**

To achieve efficient optical coupling between the incident photons and the meander area, we prepared a pair of oxygen-free copper blocks for packaging the devices, as shown in Figure 3. One of these blocks was a chip-mounting block and the other was a fiber-holding block. An OC-SNSPD chip was mounted on the former. The chip corners were adhered using electron wax, and the device was connected to a coaxial cable. The chip-mounting block had a through hole at the center of the chip-mounting area, in which the MU-type fiber ferrule (1.25 mm diameter) was inserted from the rear. In order to avoid direct contact between the mounted SNSPD chip and fiber ferrule during packaging (since direct contact can split the chip), the fiber ferrule was fixed to the fiber-holding block in advance by using an adhesive so that the distance between the end of the fiber ferrule and the rear surface of the OC-SNSPD chip, $L_{air}$, was 30 (20) μm at room (low) temperature. Since the rear surface of the OC-SNSPD chip was fixed firmly on the chip-mounting block, $L_{air}$ did not change, regardless of the substrate thickness. The fiber-holding block was joined to the chip-mounting block from the rear, as shown in Figure 3 (a), and the two blocks were accurately aligned so that the incident light spot illuminated the center of the meander area; the alignment was ensured by monitoring the device and light spot from the front side of the chip. Consequently, the distance between the fiber end and the nanowire area of the device was equal to the sum of the thickness of the substrate $L_{MgO}$ and $L_{air}$. Since the package shown in Figure 3 (b) are almost of the same size (15 mm wide, 15 mm long, and 10 mm thick) as that of the package for traditional SNSPD chips [8], they were installed in the multichannel Gifford McMahon (GM) cryocooler system without any modification.

 The cryocooler system can simultaneously cool a maximum of six SNSPD packages to 2.9 K and has a thermal fluctuation range of 10 mK [8]. Semi-rigid brass coaxial cables were introduced from each package to the output signal ports at room temperature in order to bias the current and read the output signals. The output signal port was connected to a bias tee and two low noise amplifiers (LNAs) through a coaxial cable at room temperature. The device was current biased via the dc arm of the bias tee, and the output signal was counted by pulse counter through the ac arm of the bias tee and two LNAs. Standard single mode (SM) optical fibers for telecommunication wavelengths were used for each package to input single photons. 1550- and 1310-nm-wavelength continuous laser diodes were used as the input photon sources, and they were heavily attenuated so that the photon flux at the optical input of the cryostat was $10^6$–$10^7$ photons/s. A polarization controller was inserted in front of the cryocooler optical input to control the polarization properties of the incident photons so that they matched the optimal polarization for each device (thus maximizing the DE). The system DE was defined as the output count rate divided by the input rate of photon flux to the system.

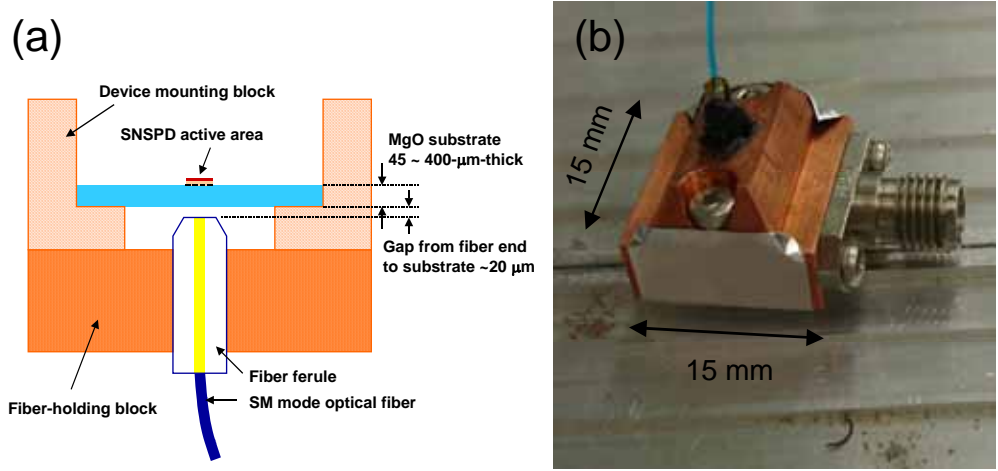

Fig. 3. (a) Schematic view of and (b) Photograph of fiber-coupled package for an OC-SNSPD chip.

## 4. Optimization for efficient optical coupling

As mentioned in section 2, reducing the substrate thickness is an effective method to decrease the distance between the fiber end and the active area of the device and achieve high optical coupling in the fiber-coupled package. To verify this assertion, the dependence of the system DE on the distance between the fiber end and nanowire area of the device was investigated by varying the substrate thickness. We measured the system DE at a dark-count rate of 100 Hz for four thicknesses of the substrate in an OC-SNSPD chip; the thickness was initially 400 μm and was then gradually reduced to 200, 100, and 45 μm. The nanowire area of the device, nanowire thickness, nanowire width, and nanowire pitch were $15 \times 15$ μm$^2$, 5.0 nm, 100 nm, and 200 nm, respectively. Since the purpose of this measurement was only to determine the relative relationship between system DE and the substrate thickness in the same device, for simplicity, an AR layer was not coated. We measured the critical current $I_c$, device resistance at 20 K, $R_{20K}$, and transition temperature $T_c$ of the device at each substrate thickness to check the degradation by thinning process. No change was observed in $R_{20K}$ and $T_c$, and the observed change in the critical current $I_c$ was 0.2%, indicating that there was no degradation by thinning process and the device DE scarcely changed before and after thinning.

The substrate thicknesses were determined by using a method originating from swept-frequency laser interferometry reported in [11]. The observation of the back-reflected power as a function of the incident light wavelength reveals periodic fringes resulting from interference between the two boundary surfaces at which optical reflection occurs. The distance $d$ between these two boundary surfaces can be estimated as the expression $d = \lambda^2/2n\Delta\lambda$, where n is refractive index, $\lambda$ is the average wavelength over the scan and $\Delta\lambda$ is the fringe spacing. In the case of the packaged OC-SNSPD, reflection of the incident light occurs at the fiber end as well as the rear and front surfaces of the substrate. Therefore, two clear periods in a fringe that originate from three surfaces should be observed. Figure 4 shows the back-reflection power as a function of the incident light wavelength for the packaged OC-SNSPD for different substrate thicknesses. Two clear fringes that originated from $L_{MgO}$ and $L_{air}$ can be seen, indicating the substrate thickness. Since it was difficult to distinguish these two fringes as the substrate thickness decreased, we also observed the substrate thicknesses directly on optical micrographs(Figure 2 (b)).

Figure 5 shows the relative system DE normalized by a value for a 400-μm substrate thickness as a function of the optical path length $L_{opt}$ between the fiber end and the device; $L_{opt}$ is defined as $L_{air} + n_{MbO}L_{MgO}$. Here, $L_{air}$ (=20 μm) is the distance between the fiber end and the rear surface of the substrate, $L_{MgO}$ is the substrate thickness, and $n_{MgO}$ (=1.7) is the refractive index of the MgO substrate. As shown in the figure, the system DE increases drastically as the substrate thickness is reduced, and the DE at a thickness of 45 μm is approximately four times higher than that at 400 μm. It is obvious that this increase is due to the improvement in the optical coupling between the incident light and the active area, achieved by reducing the substrate thickness.

We also estimated the optical coupling efficiency between the incident light and the meander area as a function of the optical path length in order to further analyze the obtained relative system DE values. At a distance $x$ from the fiber end, the light spot radius $\omega(x)$ is given by the following formula if the incident light from the fiber end is a Gaussian beam:

$$\omega^2(x) = \omega_0^2[1 + (\lambda x / \pi \omega_0^2)^2] \quad (1)$$

Here, $2\omega_0$ is the initial ($x = 0$) mode field diameter (MFD) of the optical spot, and $\lambda$ is the incident light wavelength. The power density at a distance $x$ from the center of a centered circular area with diameter $r$ is given by

$$P(r,x) = P(\infty,x)\{1 - \exp[-2r^2 / \omega^2(x)]\} \quad (2)$$

We calculated the power density within a 15 × 15 μm² nanowire area using the following parameters: $2\omega_0$ = 10.6 μm, which is the typical MFD of an SM fiber at 1550 nm, and $\lambda$ = 1550 nm. The estimated optical coupling efficiency is also shown in Figure 5 as a solid line. The dependence of the optical coupling on the optical path length agreed well with the above calculation, indicating that the relation between the measured relative system DE and $L_{opt}$ is mainly determined by the optical coupling efficiency. Further, a reduction in the substrate thickness was effective in achieving both high optical coupling and high system DE, as expected. In addition, these results indicate that an optical coupling efficiency of ~98% can be achieved at a substrate thickness of just 45 μm, and hence, it is not necessary to reduce the substrate thickness further for 15 × 15 μm² devices.

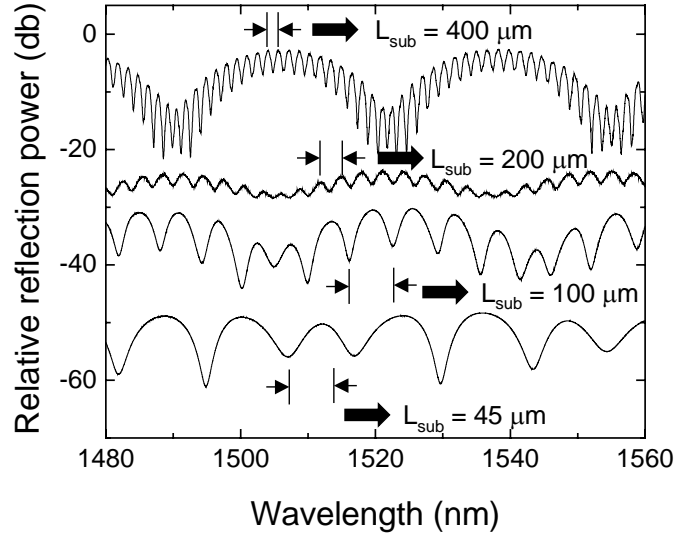

Fig. 4. Back-reflection power as a function of the incident light wavelength for a packaged OC-SNSPD for different substrate thicknesses.

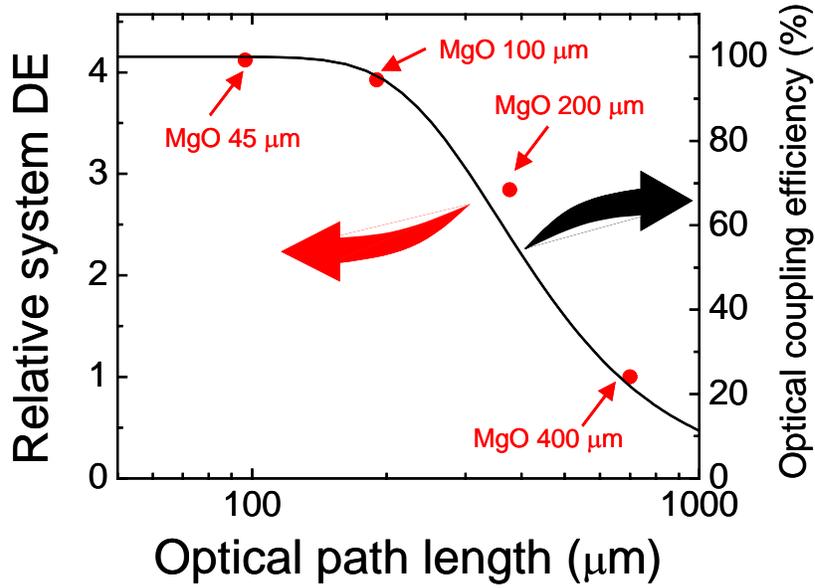

Fig. 5. Normalized relative system DE for a 400-μm substrate thickness as a function of the optical path length. The solid line shows the optical coupling efficiency calculated from a Gaussian beam approximation.

## 5. Performance evaluation

To examine the effect of integrating an optical cavity structure, we measured the system DE of an SNSPD device as a function of the dark-count rate at wavelengths of 1550 nm and 1310 nm before and after integrating an optical cavity and applying an AR layer, as shown in Figures 6 (a) and (b). Before the optical cavity was coated, the system DEs at a 100 c/s dark-count rate of the device were 2.5% and 5.0% at wavelengths of 1550 and 1310 nm, respectively. These values increased to 8.3% and 25% after integrating the optical cavity, even though the optical cavity was specially designed for a wavelength of 1550 nm. The values exceeded 15% and 30% in the region with a high dark count rate, where the bias current was just below $I_c$ (~0.99$I_c$). On the other hand, at 1550 nm, the system DE increased slightly to 9.5% (dark-count rate: 100 c/s), while at 1310 nm, it decreased to 18%. These results indicate that the presence of the AR layer for a wavelength of 1550 nm favorably affects the sensitivity at 1550 nm, but degrades the sensitivity at a wavelength of 1310 nm. The enhancement factor , which is the ratio of the system DE before integrating the optical cavity to that after, was 3.8 at 1550 nm and 5.0 at 1310 nm. These values were greater than expected because the absorptance of 4-nm-thick NbN films measured by a spectrophotometer was 32% around 1550 nm, indicating that the maximum enhancement factor should not significantly exceed ~3. This discrepancy might be due to the differences of absorptance between patterned nanowire and unpatterned film. The interferences in OC-SNSPD packaging as described above can be also considered to cause this discrepancy. The interferences limit the sensitivity, resulting in variations of enhancement factor. A systematic investigation of the dependence of DE on the wavelength is necessary to clarify the relation between the interferences and system DE and to obtain the accurate enhancement degree for the optical cavity and AR coating; Such an investigation will be performed in a future study.

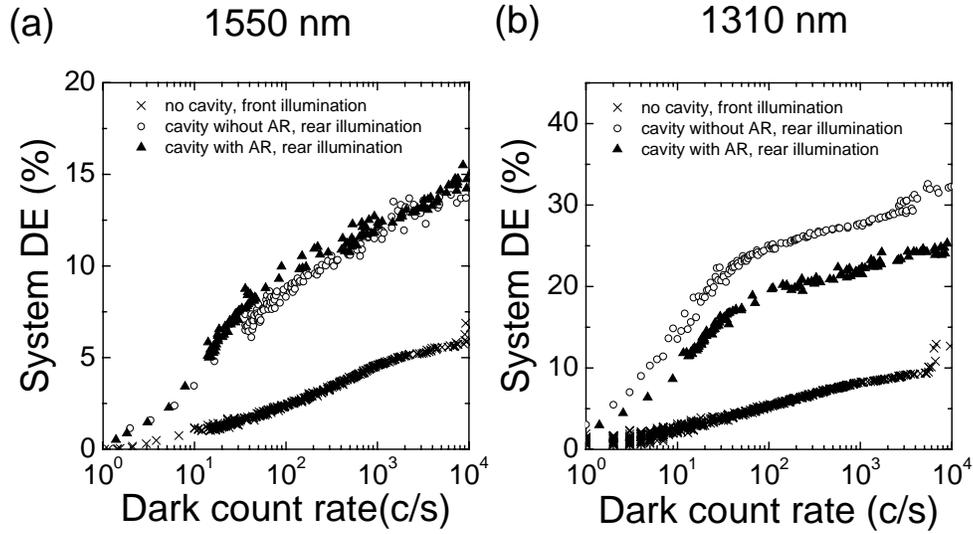

Fig. 6. System DE of a device as a function of the dark-count rate at wavelengths of (a)1550 nm and (b) 1310 nm, before and after integrating an optical cavity and an AR coating.

**6. Conclusion**

We have successfully fabricated OC-SNSPD devices and developed a compact packaging technique for establishing efficient optical coupling between the OC-SNSPD device and an optical fiber. The packaged devices can be easily installed in a practical multichannel system. System DE measurement and estimation from the Gaussian beam approximation indicated an improvement in the optical coupling efficiency up to 98% when the substrate thickness was reduced to 45 μm. The obtained system DE was higher than that of SNSPD devices without an optical cavity by more than a factor of three and exceeded 9.5% and 25% at wavelengths of 1550 nm and 1310 nm, respectively (nominal dark-count rate: 100 c/s). This practical multichannel system with high system DE is expected to lead to a significant improvement in QKD and other quantum information processing experiments.